\documentclass[12pt,preprint]{aastex}

\newcommand{\etal}{et~al.\ }
\newcommand{\HI}{\hbox{{\rm H}\kern 0.1em{\sc i}}}
\newcommand{\HII}{\hbox{{\rm H}\kern 0.1em{\sc ii}}}
\newcommand{\NII}{\hbox{[{\rm N}\kern 0.1em{\sc ii}]}}
\newcommand{\Ha}{\hbox{{\rm H}\kern 0.1em$\alpha$}}
\newcommand{\msun}{${\rm M}_{\odot}$}

\begin{document}
\slugcomment{The Astrophysical Journal Letters, accepted 2012 February 24}

\shortauthors{KNIERMAN ET~AL.}
\shorttitle{Unexpected Star Formation in NGC 2782}

\title{Tidal Tails of Minor Mergers: Star Formation Efficiency in the Western Tail of NGC 2782}

\author{Karen Knierman}
\affil{School of Earth \& Space Exploration, Arizona State University,
550 E. Tyler Mall, Room PSF-686 (P.O. Box 871404),
Tempe, AZ 85287-1404}
\email{karen.knierman@asu.edu}

\author{Patricia M. Knezek}
\affil{WIYN Consortium, Inc., 
950 North Cherry Avenue,
Tucson, AZ  85719}
\email{pknezek@noao.edu}

\author{Paul Scowen}
\affil{School of Earth \& Space Exploration, Arizona State University,
550 E. Tyler Mall, Room PSF-686 (P.O. Box 871404),
Tempe, AZ 85287-1404}
\email{paul.scowen@asu.edu}

\author{Rolf A. Jansen\altaffilmark{1}}
\affil{School of Earth \& Space Exploration, Arizona State University,
550 E. Tyler Mall, Room PSF-686 (P.O. Box 871404),
Tempe, AZ 85287-1404}
\email{rolf.jansen@asu.edu}

\author{Elizabeth Wehner}
\affil{Department of Astronomy, Haverford College, Haverford, PA 19041}
\email{ewehner@haverford.edu}

\altaffiltext{1}{Department of Physics, 550 E. Tyler Mall, Room PSF-470 (P.O.Box 871504), Tempe, AZ 85287-1504}

\begin{abstract}

While major mergers and their tidal debris are well studied, they are less common than minor mergers (mass ratios $\la0.3$).  
The peculiar spiral NGC 2782 is the
result of a merger between two disk galaxies with a mass ratio of $\sim4:1$
occurring $\sim200$ Myr ago. This merger produced a molecular and {\HI} rich,
optically bright Eastern tail and an {\HI}-rich, optically faint
Western tail.   Non-detection of CO in the Western Tail by \citet{braine} suggested that
star formation had not yet begun to occur in that tidal tail.  However, deep {\Ha} narrowband images 
show evidence of recent star formation in the Western tail.  Across the entire Western tail,
we find the global star formation rate per unit area ($\Sigma_{\rm SFR}$) several orders of magnitude less than expected from the total gas density.  Together with extended FUV+NUV emission from \emph{Galaxy Evolution Explorer} along the tail, this indicates a low global star formation efficiency in the tidal tail producing lower mass star clusters.  
The {\HII} region we observed has a \emph{local} (few-kpc scale)
$ \Sigma_{\rm SFR}$ from {\Ha} that is less than that expected from the
 \emph{total} gas density, which is consistent with other observations
 of tidal debris.
The star formation efficiency of this {\HII} region inferred
 from the total gas density is low, but normal when inferred from the
 molecular gas density.
These results suggest the presence of a very small, locally dense region in the Western tail of NGC 2782 or of a low metallicity and/or low pressure star forming region.

 \end{abstract}


\section{Introduction}
Major mergers of spiral galaxies  are known to create
structures such as tidal dwarf galaxies (TDGs) and star clusters
within their debris \citep[e.g.,][]{duc,wfdf02,knierman,mullan}.  While examples of major mergers are well known \citep[e.g., NGC
4038/9 ``The Antennae'';][]{whitmore99}, interactions between equal
mass galaxies are relatively rare compared to minor mergers
(mass ratios of $\la
0.3$).  As part of a larger study, we aim to
understand how these frequent encounters shape galactic structure and  
probe star formation in gas that may be marginally stable. 
Previous work studied how neutral hydrogen may affect
star cluster formation in tidal debris \citep{aparna, mullan}, but studies of 
molecular gas in tidal debris have focused on larger episodes of star formation such
as those resulting in the formation of TDGs \citep{braine}.  This work examines star formation on smaller scales in the tidal debris of the minor merger NGC 2782.  

NGC 2782, a peculiar spiral at a distance of ($39.5 \pm 2.8$) Mpc\footnotemark[1],  is undergoing a
nuclear starburst \citep{devereux}.  \citet{smith94} used a restricted three-body dynamical model  to show that NGC 2782 is the result
of a merger of two disk galaxies
with a mass ratio of $\sim0.25$ occurring $\sim200$ Myr ago.  It has two tidal tails: an Eastern tail which
has a concentration of {\HI} and CO at its base and a gas-poor
optically bright knot $2.7'$ from the center; and an {\HI}-rich, optically faint
Western tail  \citep{smith94}.  \citet{mullan} in their $V$ and $I$ band \emph{Hubble Space Telescope}/WFPC2 survey of tidal tails find 87 star cluster candidates in the Eastern tail of NGC 2782 and 10 candidates in the Western tail. 

Non-detection of CO at the
location of {\HI} knots in the Western tail led \citet{braine} to
argue that the {\HI} ``has presumably
not had time to condense into H$_2$ and for star formation to begin."
However, if this tail was pulled from the lower metallicity outer regions of the spiral
galaxy like TDGs \citep{duc} or the merged dwarf galaxy, the lower metallicity may affect the conversion factor between CO and H$_2$ and result in an underestimated molecular mass.  It is possible for  H$_2$ to be present despite CO being undetected.  While the blue colors in the Western tail suggest that it formed from the disruption of the dwarf companion, the $m_{\rm HI}/L_B$ ratios suggest that some gas must have originated in NGC 2782's gaseous disk and is therefore mixed composition \citep{wehnerthesis}.  

\footnotetext[1]{From NED, corrected for Virgo, Great Attractor, and Shapley, which we will use for the duration of this paper. \citet{smith94,braine} use distances of 34 Mpc and 33 Mpc, respectively.}

We obtained new {\Ha} observations to determine the star formation efficiency
in the Western tail of NGC 2782.  Section~2 presents observations, calibration, and results.  In Section 3, we discuss global and local star formation in the tail and relate it to star formation in general.  

\section{Observations}
\begin{figure}
\includegraphics[width=\textwidth]{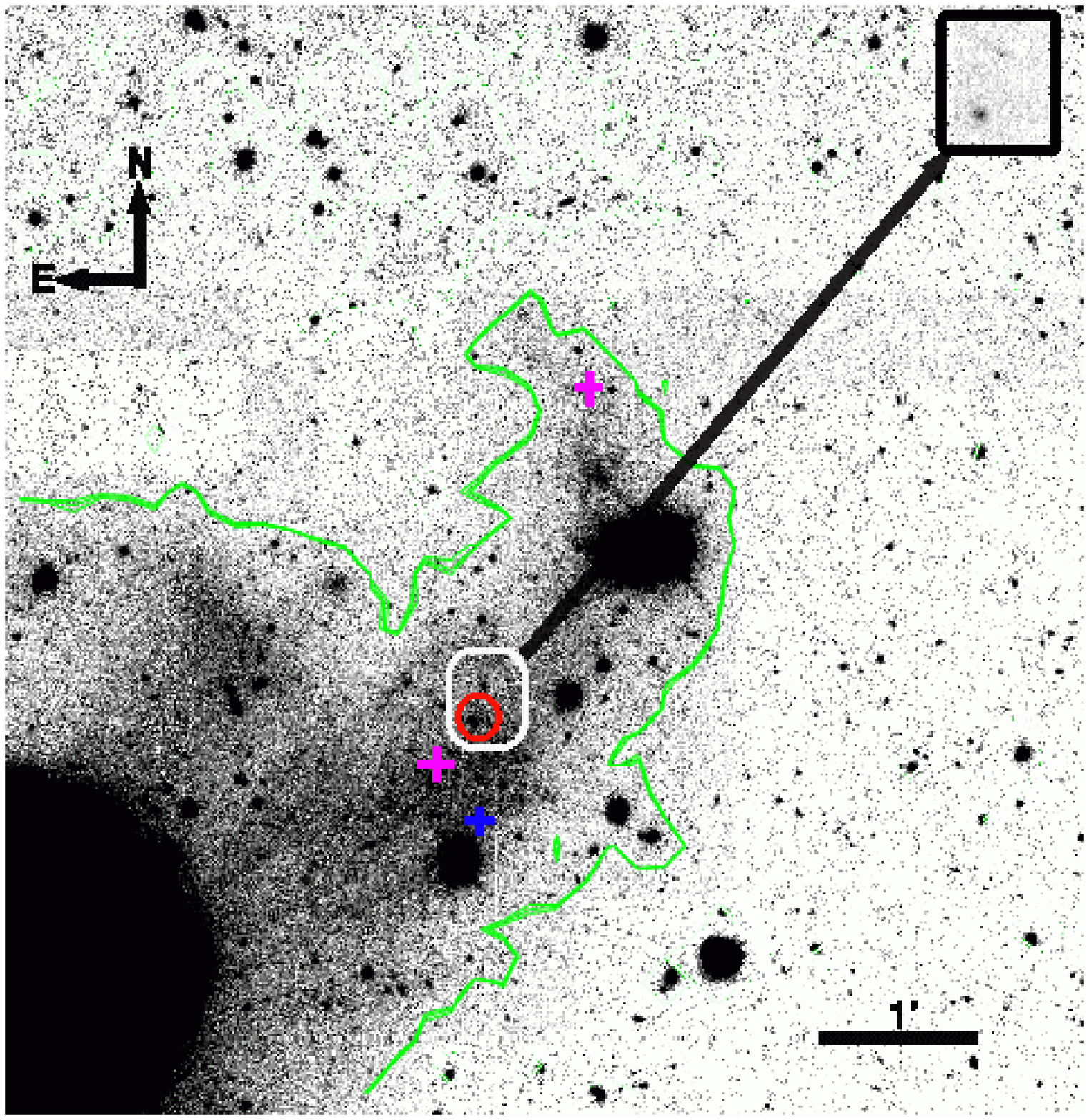}
\caption{\emph{V} image of Western tail of NGC 2782.  
The green contour indicates the area defined as the Western Tidal Tail.
Red circle marks the location of the {\Ha} source. The crosses mark the locations of massive {\HI} clouds\citep{smith94}.  Magenta crosses mark the
locations of CO observations from \citet{braine} for the north location and \citet{smith99} for the south location. The inset image is the continuum subtracted {\Ha} image of area indicated in the white box. }
\label{fig:VimageW}
\end{figure}

Images in \emph{UBVR} and {\Ha} were taken with the Loral 2K CCD imager at the Lennon 1.8m Vatican
Advanced Technology Telescope (VATT) (6.4$'$ field of view, 0.375{\arcsec} pixel$^{-1}$). 
{\Ha} images (6$\times$1200\,s) used an 88\,mm
Andover three-cavity interference filter ($\lambda_c$=6630\AA; FWHM=70\AA).
We observed the tail in Kron-Cousins \emph{R} (3$\times$300\,s) to
 allow continuum subtraction, following \citet{lee}. 
Images were reduced using standard IRAF\footnote{IRAF is distributed by the National Optical Astronomy
Observatory, which is operated by the Association of Universities for Research in Astronomy, Inc., under cooperative agreement with the National Science Foundation.} tasks.  
The inset in Figure~\ref{fig:VimageW} shows the continuum-subtracted {\Ha} image that contained the only H$\alpha$
emission-line source detected at more than $10\sigma$ in the Western tail.

\subsection{{\Ha} calibration}
We calibrated our {\Ha} images using observations of 3--5
spectrophotometric standard stars from \citep{okestone}.  Zeropoints were
obtained by comparing the integral over the filter response function
of their spectral energy distribution and the instrumental magnitude
from aperture photometry.  Extinction corrections assumed a standard
atmospheric extinction coefficient of 0.08 mag\,airmass$^{-1}$ \citep{lee}.  
The dispersion of the zeropoints from individual standard
  stars was typically 0.02\,mag.

Following \citet{lee}, we removed the contribution to the {\Ha} flux of 
 the {\NII} doublet ($\lambda$6548,6583) and emission line flux  from the  $R$ filter.  
 We used an empirical relation between
metallicity and the \NII$/${\Ha} ratio from Figure 9 of
\citet{vanZee98}.
For a metallicity of 0.4Z$_{\odot}$, $12 + \log (\rm O/H)
= 8.06$ gives $\log ($\NII $/$\Ha$) = -1.3$, from which follows the {\Ha} flux.

\subsection{Results}
The {\Ha} observations yielded a detection of one source in the
Western tail centered on $\alpha=$ 9:13:51.2, $\delta=$+40:08:07 (see
Figure \ref{fig:VimageW})
with L$_{\rm H\alpha} = (1.9\pm 0.3) \times 10^{39}$ erg s$^{-1}$.  
For comparison, this {\HII} region is 
 fainter than the massive star cluster 30 Dor (L$_{\rm H\alpha} = 6
\times 10^{39}$ erg s$^{-1}$), but $>1000$ times brighter than Orion with its handful of O-stars (L$_{\rm H\alpha} =
10^{36}$ erg s$^{-1}$).  It is consistent with the formation of a large star cluster. 
This {\HII} region has also been detected by \citet{bournaud} and, recently, by \citet{werk}. The {\HII} region is located $\sim20$\arcsec  \ away, but well within the 55\arcsec \ half power beam size, from the location where \citet{smith99} searched for CO(1-0).

\section{Discussion}

We compare the star formation rate (SFR) per unit area  ($\Sigma_{\rm SFR}$) from {\Ha} to that obtained from the observed gas density using the Kennicutt law, and the SFE in the tail to that seen in other tidal debris, normal galaxies, and starbursts. $\Sigma_{\rm SFR}$ from {\Ha} for the whole tail is much less than expected given the observed gas density.  With only one {\Ha} region in the tail, the derived $\Sigma_{\rm SFR}$ is a lower limit, as most of the stars forming are late B and A stars based on ultraviolet emission.  This indicates that there is a lower SFE in the tail resulting in the formation of fewer high mass stars.  Star formation on the few-kpc scale represents a $\Sigma_{\rm SFR}$ that is less than expected from the Kennicutt law, using the total gas surface density and the observed {\Ha}.  Since the original Kennicutt law was formulated using observations of spiral disks, this indicates that the star formation in the tail is less efficient than in spiral disks.  Using the molecular gas depletion time, the SFE of the {\HII} region is similar to the tidal debris regions of Arp 158 and normal galaxies but lower than observed in starburst galaxies.  Using only the {\HI} gas as a tracer of the available material, the SFE is higher than seen in the outer disks of spiral galaxies.  Given a low SFE from the total gas and a normal SFE from the molecular gas, the observed {\HII} region may be a very small, locally dense region.  The lack of observed CO emission could be due to destruction of molecular gas by FUV, effects of beam dilution, the influence of low metallicity on the CO-H$_2$ conversion factor, or a low pressure gas environment.

The SFR from {\Ha} is  \citep[Equation (2) in][]{kennicutt}: 
SFR (\msun yr$^{-1}) = 7.9\times10^{-42} \rm L$(\Ha) (ergs  s$^{-1})$.  The expected SFR from the gas
surface density is  \citep[Equation (7) in][]{kennicutt}:  

\begin{equation}
\Sigma_{\rm SFR} (\rm gas) = 2.5 \times 10^{-4} (\frac{\Sigma_{\rm gas}}{1 \rm M_{\odot} pc^{-2}})^{1.4} \rm M_{\odot} yr^{-1} kpc^{-2}
\end{equation}.

Table~\ref{sfrcomp} compares the global and local SFR in the Western tail:  tail location, area of {\HI} clump or entire tail, {\Ha} SFR with error,
SFR per unit area from {\Ha} ($\Sigma_{\rm SFR}(\rm H\alpha)$) with error, mass of {\HI}, mass of molecular gas, total gas surface density, and SFR per unit area from gas density ($\Sigma_{\rm SFR}$ (gas)).  

\begin{deluxetable}{cccccccc}
\tabletypesize{\scriptsize}
\setlength{\oddsidemargin}{-0.5in}
\tablecaption{Comparison of Star Formation Rates in Western Tail of NGC 2782\label{sfrcomp}}
\tablewidth{7.15in}
\tablewidth{0pt}
\tablehead{\colhead{Location} & \colhead{Area} & \colhead{H$\alpha$ SFR} & \colhead{$\Sigma_{\rm SFR}(\rm H\alpha)$} & \colhead{M$_{\HI}$\tablenotemark{a}} & \colhead{M$_{\rm mol}$\tablenotemark{b}} & \colhead{$\Sigma_{\rm gas}$\tablenotemark{c}} & \colhead{$\Sigma_{\rm SFR}$(gas)\tablenotemark{d}}  \\
\colhead{} & \colhead{(${\rm kpc^2}$)} & \colhead{($M_\odot{\rm yr^{-1}}$)} & \colhead{($M_\odot{\rm yr^{-1} kpc^{-2}}$)} & \colhead{($10^8 M_\odot$)} & \colhead{($10^8 M_\odot$)} & \colhead{($M_\odot{\rm pc^{-2}}$)} & \colhead{($M_\odot{\rm yr^{-1} kpc^{-2}}$)}}
\startdata

\HI-N & 8.6 & $<0.0003$ & $<0.00003$ & $0.73$ & $<0.086$\tablenotemark{e} & $<12.9$ & $<0.006$ \\
\HI-mid & 14.7 & 0.015(.002) & 0.001(0.0002) & $1.15$ & $<0.16$\tablenotemark{f} & $<12.2$ & $<0.005$ \\
\HI-S & 19.3 & $<0.0003$ & $<0.00002$ & $1.16$ & $<0.16$\tablenotemark{f} & $<9.4$ & $<0.004$ \\
\hline \\
W Tail &  2300 & 0.015 & 0.000009 & 20 & $<0.4$\tablenotemark{e,f} & $<11.7$ & $<0.005$ \\
 
\enddata

(a) \citet{smith94}, corrected for distance,  (b) M$_{\rm mol}$ inferred from CO observations,  (c) Includes helium (M$_{\rm gas} = 1.36(\rm M_{\HI} + M_{\rm H_2}$) (d) From \cite{kennicutt},  $\Sigma_{\rm gas}$ includes only {\HI} and H$_2$ (e) \citet{braine}, corrected for distance, (f) \citet{smith99}, corrected for distance

\end{deluxetable}

\subsection{Star Formation on Global Scales}
Using the entire area of Western tail of NGC 2782, the global $\Sigma_{\rm SFR} (\Ha) = 9 \times 10^{-6}$ \msun yr$^{-1}$kpc$^{-2}$ is three
orders of magnitude below the expected $\Sigma_{\rm SFR} (\rm gas) < 5 \times 10^{-3}$ \msun yr$^{-1}$kpc$^{-2}$.
The $\Sigma_{\rm SFR} (\Ha)$ is also three orders of magnitude lower than those typical for spiral and dwarf galaxies, but there is a significant dilution factor due to the large area of the tidal tail and low density of stars and gas therein.

The Magellanic Stream
is a local example of a gas tail of presumed tidal origin with no 
star formation.  \citet{putman} measured the total {\HI} gas mass
in the Stream to be $2.1 \times 10^8$ \msun.  By converting the angular size of the Stream
($100 ^{\circ} \times 10 ^{\circ}$) to projected physical size using a distance of 55 kpc \citep{putman}, we 
infer an area of 940.9 kpc$^2$.  Using the resulting gas surface density of 
$\Sigma_{\rm \HI} =  2.2  \times 10^5$ \msun kpc$^{-2}$,
the Magellanic Stream has an expected $\Sigma_{\rm SFR} = 3 \times
10^{-5}$ \msun yr$^{-1}$ kpc$^{-2}$, two orders of magnitude lower than the Western tail of NGC 2782.  

\begin{figure}
\includegraphics[width=\textwidth]{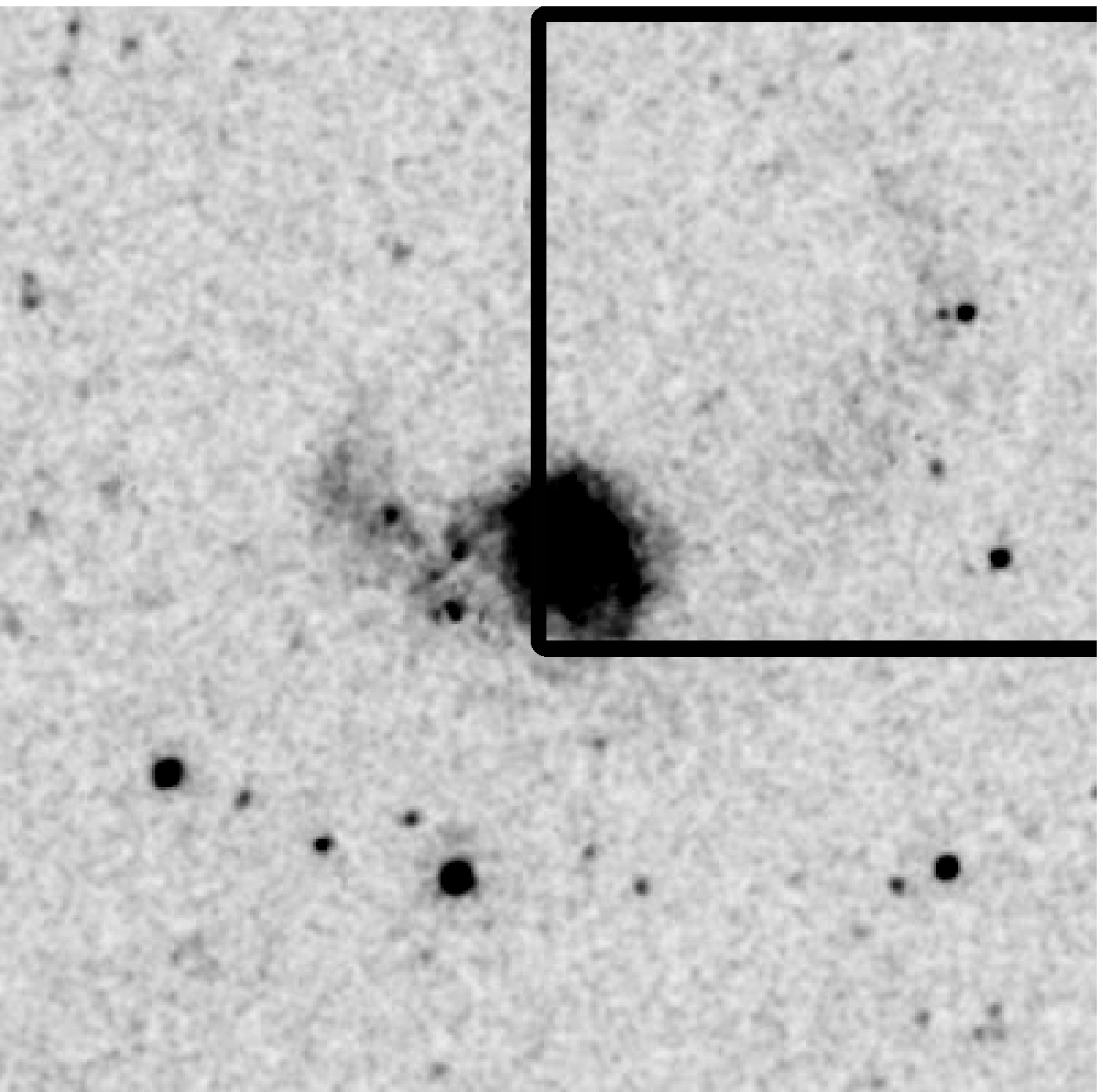}
\caption{Galex composite image of NGC 2782.  The box indicates the region covered by the optical image of Fig.~\ref{fig:VimageW}}
\label{fig:Galex}
\end{figure}

The difference between the $\Sigma_{\rm SFR}$ values 
indicates a lower SFE.  However, the SFR is a lower limit since 
there is one {\HII} region and {\Ha} represents star formation in the last 5 Myr.  Also the
SFR in the tail could be higher than inferred from {\Ha} if it is 
predominately of a Taurus-Auriga type \citep{kenyon}, producing few star clusters with high
mass stars.  If so, the color would be blue, but no {\Ha} would be observed.  
To examine this further, FUV and NUV images of NGC 2782 from the
\emph{Galaxy Evolution Explorer} (\emph{GALEX}) All-sky Imaging Survey \citep[AIS;][]{galex} were inspected.  
As seen in Figure~\ref{fig:Galex}, faint UV emission is detected along the Western tidal tail, indicating
the presence of young stellar populations, likely dominated by B and A stars.  Since there is no {\Ha} 
emission (except for the single knot) the star clusters forming along the tail were likely of low mass and
had a negligible probability of forming early B and O stars.  

\subsection{Star Formation on Local Scales}
We also examine star formation within Western tail {\HI} regions.
\citet{smith94} measured 10 massive {\HI} clumps with masses from 
$3\times10^7$ {\msun} to $1.8\times10^8$
\msun.  Only three of these {\HI} clumps have star cluster
candidates \citep{mullan}.  Since only one {\HII} region was
found in the Western tail, we use the {\Ha} detection limit as a limit
for the high-mass SFR for the other two regions.  The
crosses in Figure \ref{fig:VimageW} show the location of
these {\HI} clumps with their SFR in Table~\ref{sfrcomp}.  

The star formation on few-kpc scales associated with the {\HII} region is lower than expected from the Kennicutt law.  
$\Sigma_{\rm SFR}$ calculated from the {\Ha}
luminosity is $ 0.001 \pm 0.0002$ $M_\odot{\rm yr^{-1} kpc^{-2}}$.  
Due to the non-detection of CO in the Western
tail and using a standard CO to H$_2$ conversion factor, we are only able to establish upper limits to the SFR from the gas density of  $\Sigma_{\rm SFR}(\rm gas) < 0.005$ \msun yr$^{-1}$ kpc$^{-2}$.
While follow-up spectroscopy is necessary to
determine the composition of the gas in the Western tail of NGC 2782,
 TDGs in major mergers are shown
to have $\sim0.3Z_{\odot}$ 
\citep{duc}.   The standard CO to H$_2$ conversion factor 
($\alpha_{\rm CO 1-0} = 4.3 $  \msun (K km s$^{-1}$ pc$^2)^{-1}$ ) is based on observations of the 
Milky Way.  
Using galaxies with $z\le1$, \citet{genzel} find a linear relation given by $\log \alpha_{\rm CO 1-0} = 12.1 - 1.3\mu_0$ where $\mu_0 = 12 + \log(\rm O/H)$.  For a metallicity of 0.3$Z_{\odot}$ (or $\mu_0 = 8.19$), the appropriate conversion factor is $\alpha_{\rm CO 1-0} = 27.5$  \msun (K km s$^{-1}$ pc$^2)^{-1}$ giving a factor of six higher molecular mass limit  
 (M$_{\rm mol} \leq 6 \times 10^7$ \msun) than that from the 
standard conversion factor (M$_{\rm mol} \leq 9 \times 10^6$ \msun).  This would then give a larger expected $\Sigma_{\rm SFR}(gas) < 0.01$ \msun yr$^{-1}$ kpc$^{-2}$ and would give an even larger difference from the measured {\Ha} SFR. 
\citet{boquien} use multiwavelength data of Arp 158 to study the local Schmidt-Kennicutt law in a merger.  They find that star forming regions in the tidal debris follow a different Schmidt-Kennicutt law than those in the central regions of the merger, falling along a line of similar slope to \citet{daddi}, but offset so that the same gas density gives lower values of SFR.  Plotting our {\HII} region in the Western tail of NGC 2782 on Figure 6 of \citet{boquien}, we find it to be consistent with quiescent star formation as seen in the tidal debris of Arp 158.   This may indicate that star formation in tidal debris is less efficient than that in the central regions of mergers and in normal galaxies.

However, this {\HII} region has a normal SFE using the molecular gas limit.  
The depletion timescale of the molecular gas is calculated by $\tau_{\rm dep, H_2} = $M$_{\rm mol}/$SFR.  For the {\HII} region in the western tail of NGC 2782, $\tau_{\rm dep} < 1$ Gyr which is comparable to the molecular gas depletion timescales determined for the star forming regions in Arp 158 \citep[$\tau_{\rm dep} \sim 0.5-2$ Gyr;][]{boquien} and in TDGs \citep[$\tau_{\rm dep} \sim 0.8-4$ Gyr;][]{braine}.  These ranges are also similar to the average gas depletion timescales in spiral galaxies. 
The inverse of $\tau_{\rm dep}$ is related to the star formation efficiency.  
 For the {\HII} region in the western tail of NGC 2782, SFE$=(\tau_{\rm dep})^{-1} > 9.3\times10^{-10}$ yr$^{-1}$.  In contrast to the low SFE implied from the total gas density, this {\HII} region appears to have a similar SFE to the tidal tail regions in Arp 158 and in normal spiral galaxies based on the molecular gas upper limit but lower than dense starburst nuclei \citep[e.g., NE region in Arp 158;][]{boquien}.   \citet{bigiel} find very low SFE ($< 9\times10^{-11}$ yr$^{-1}$) in the outer disks of spiral galaxies using FUV and {\HI} observations.  The {\HII} region in the western tail of NGC 2782 has a higher SFE than these outer disk regions ($(\tau_{\rm dep,{\rm \HI}})^{-1} = 1.3\times10^{-10}$ yr$^{-1}$) using only the {\HI} gas mass.

Since there is a low SFE using the total gas density and a normal SFE using the molecular gas limit, this {\HII} region may be very small and dense or something else entirely.  
Due to the lack of wide-spread \emph{massive} star formation, using {\Ha} as the star formation indicator likely underestimates the true nature of star formation in the Western tidal tail.  This means that our SFE estimates are lower limits, particularly when combined with the upper limit on the molecular gas mass.  The discrepancies between the SFE may indicate that there is a denser region of star forming gas that is too small to have been observed.  Unlike the very dense regions in the central regions of mergers such as those in the models of \citet{teyssier}, there may still be elevated levels of star formation across mergers even out in the tidal debris regions.  
Star formation in mergers likely depends on local conditions at a scale of 1 kpc, which is the size of the gravitational instabilities in the ISM of mergers and the injection scale of turbulence \citep{elmegreen93, elmegreen04}.

\subsection{Impact on Star Formation}
Tidal tails provide laboratories for star formation under conditions
very different from quiescent galaxy disks. With low gas pressures and
densities and small amounts of stable molecular gas they are perhaps
at the edge of the parameter space open to star formation.
The Western tail of NGC 2782 is {\HI} gas rich, but CO is not
observed in the massive {\HI} knots in the tail. 
This study finds an {\HII} region in the
tidal tail indicating recent star formation. Clearly, the lack of {\it observable} CO
does not guarantee the absence of recent star formation. 
The presence of a young star cluster in a tail without detectable
molecular gas requires one of two situations; either there is no 
CO, or it escapes detection at the sensitivity of current instrumentation. 

If the molecular gas is absent, it may
be because it is short-lived. 
This is most likely the result of a strong ambient FUV radiation field
produced by the high local SFR. However, the Western
tail does not have a high SFR, so this is unlikely to cause
the lack of observed molecular gas.

The molecular cloud may be too small to be observed.  
In general, H$_2$ is not directly detectable, so we must rely on surrogate tracers such as CO 
\citep[][and references therein]{solomon}. At the
distance of NGC 2782, an arcsecond corresponds to a physical scale of 190 pc. 
This is not an unusual size for molecular
clouds in the Galaxy; compact clouds may be smaller still. The IRAM observations of
\citet{braine} had a 21\arcsec \ half power beam size for their CO(1-0) observations while the Kitt Peak 12m observations of \citet{smith99} had a 55\arcsec \ half power beam size.  If there are only 
one or a few clouds at the location of their observations, then beam dilution is 
a major detriment to the detection of CO at the {\HII} region and in the massive {\HI} clouds.  

Physics also works against the
detection of molecular gas. If the gas is drawn from the
dwarf or the outer regions of the large galaxy, it may be
deficient in heavy elements. The CO to H$_2$ conversion factor 
can be different for lower metallicities, meaning a larger H$_2$ mass for a given CO flux. 
Also, CO does not form in a molecular cloud until
$A_V \ge 3$, while H$_2$ forms at $A_V < 1$ \citep{ht97}. In a
low pressure environment such as low gas density tidal debris, a substantial amount of molecular gas
can exist at low $A_V$ that will not be detectable through CO.  
Theoretical models \citep{wolfire} show that the fraction of molecular mass in the  ``dark gas" (H$_2$ and C$^+$) is $f \sim 0.3$ for typical galactic molecular clouds, increasing for lower $A_V$ and lower metallicity.

\section{Conclusions}
While the molecular gas rich Eastern tail of NGC 2782 was known to form stars, we report the detection of recent star formation in the {\HI} rich but molecular gas poor Western tail. 
This is contrary to the conclusion of \citet{braine} that the
lack of detected molecular gas in the Western tail implies that no stars are forming there.  
Globally, we find that $\Sigma_{\rm SFR}$ based on our {\Ha} observations is several orders of magnitude less than expected from the ${\rm \HI + H_2}$ gas density.  {\Ha} observations provide only a lower limit on current SFRs, as \emph{GALEX} observations show extended FUV+NUV emission along the tail. This indicates star formation is less efficient across the tail, forming lower mass star
 clusters.
We find that the observed \emph{local} $\Sigma_{\rm SFR}$ from {\Ha} is $\sim$20\% of that expected from the local total gas density, consistent with that observed in the tidal debris of Arp 158.  The {\HII} region has a low SFE considering the \emph{total} gas density, but a normal SFE considering the low molecular gas density.  This {\HII} region in the Western tail of NGC 2782 may be a very small, dense region the molecular gas in which is not observable with current instruments or may be indicative of star formation in low metallicity and/or low pressure regimes.

\begin{acknowledgments}
We acknowledge helpful comments from the referee, discussions with Janice Lee, Shoko Sakai, Jose Funes, and Jacqueline Monkiewicz, and Rob Kennicutt for use of his narrowband filter.  KK was supported by the University of Arizona/NASA Space
Grant Graduate Fellowship and through the NASA Herschel Science Center.  We used the NASA/IPAC Extragalactic Database (NED) which is operated by JPL, California Institute of Technology, under NASA contract.  This work is based on observations with the Vatican Advanced Technology Telescope (VATT): the Alice P. Lennon Telescope and the Thomas J. Bannan Astrophysics Facility.
\end{acknowledgments}

 \end{document}